%% file: main.tex
\documentclass{research4cacm}

\usepackage{colortbl}
\usepackage[dvipsnames]{xcolor}

\input{macros}

\allowdisplaybreaks

\newcommand{\outputsize}{|\problemname{output}|}
\newcommand{\FECP}{\functionname{P}}

\begin{document}
%

\title{Juggling Functions Inside a Database\thanks{The original version 
of this paper is entitled ``FAQ: Questions Asked
Frequently" and was published in the proceedings of the PODS'16 conference. 
This work was partly supported by NSF grant CCF-1319402 and by DARPA under agreement \#FA8750-15-2-0009.  The U.S.
Government is authorized to reproduce and distribute reprints for Governmental
purposes notwithstanding any copyright thereon.}}
%
%
%
%
\numberofauthors{3} 
\author{
\alignauthor
Mahmoud Abo Khamis\\
       \affaddr{LogicBlox Inc.}
	\email{\small mahmoud.abokhamis@logicblox.com}
\alignauthor
Hung Q. Ngo\\
       \affaddr{LogicBlox Inc.}
	\email{\small hung.ngo@logicblox.com}
\alignauthor 
Atri Rudra\\
       \affaddr{University at Buffalo, SUNY}
	\email{\small atri@buffalo.edu}
}

\maketitle

\input{abstract}

\input{intro}
\input{insideout}

\input{theory}

\input{practice}
\input{conclusion}

{\bibliographystyle{abbrv} 
\small\bibliography{main}}

%
\balancecolumns
\end{document}

%% file: macros.tex
\usepackage{subfig}
\usepackage{tikz,tikz-qtree,tikz-qtree-compat}
\usetikzlibrary{positioning,shapes,shadows,arrows,calc}
\usepackage{xspace}

\renewcommand{\vec}[1]{\ensuremath\boldsymbol{#1}}

\colorlet{LightViolet}{violet!40}
\colorlet{LightRed}{red!40}
\colorlet{LightOrange}{orange!40}
\colorlet{LightGreen}{green!40}
\colorlet{LightBlue}{blue!40}
\colorlet{DarkGreen}{green!50!black}
\colorlet{DarkRed}{red!70!black}
\colorlet{DarkCyan}{red!70!black}
\colorlet{DarkBlue}{blue!80!black}
{\definecolor{DarkOrange}{rgb}{.8, 0.4, 0.0}
\definecolor{Airforceblue}{rgb}{0.36, 0.54, 0.66}

\newcommand{\DarkBlue}[1]{{\color{DarkBlue} #1}}

\newcommand{\DarkGreen}[1]{{\color{DarkGreen} #1}}

\newcommand{\LB}{\text{\sf LogicBlox}}
\newcommand{\LogiQL}{\text{\sf LogiQL}}

\newcommand{\faqcs}{\text{\sf FAQ-SS}}

\newcommand{\faq}{\text{\sf FAQ}}

\newcommand{\complexityclass}{\mathbf}

\newcommand{\np}{\complexityclass{NP}}

\newcommand{\functionname}[1]{\text{\sf #1}}

\newcommand{\faqw}{\functionname{faqw}}

\newcommand{\Fss}{\functionname{$F$-ss}}
\newcommand{\htw}{\functionname{htw}}
\newcommand{\fhtw}{\functionname{fhtw}}

\newcommand{\atoms}{\functionname{atoms}}
\newcommand{\Dom}{\functionname{Dom}}

\newcommand{\vars}{\functionname{vars}}

\newcommand{\true}{\functionname{true}}
\newcommand{\false}{\functionname{false}}

\newcommand{\InsideOut}{\functionname{InsideOut}}

\newcommand{\problemname}[1]{\text{\sf #1}}
\newcommand{\mcm}{\problemname{MCM}}
\newcommand{\map}{\problemname{MAP}}

\newcommand{\csp}{\problemname{CSP}}
\newcommand{\pgm}{\problemname{PGM}}

\newcommand{\scq}{\problemname{\#CQ}}
\newcommand{\sqcq}{\problemname{\#QCQ}}

\newcommand{\EVO}{\problemname{EVO}}
\newcommand{\LE}{\problemname{LinEx}}

\newcommand{\CWE}{\problemname{CWE}}
\newcommand{\qcq}{\problemname{QCQ}}

\newcommand{\lftj}{\problemname{LFTJ}}

\newcommand{\agm}{\text{\sf AGM}\xspace}

\newcommand{\calE}{\mathcal E}

\newcommand{\calH}{\mathcal H}
\newcommand{\calV}{\mathcal V}

\newcommand{\F}{\mathbb F} 
\newcommand{\Z}{\mathbb Z} 
\newcommand{\N}{\mathbb N} 
\newcommand{\R}{\mathbb R} 
\newcommand{\D}{\mathbf D} 

\newcommand{\poly}{\textnormal{poly}}

\newcommand{\opt}{\textnormal{\sc opt}}

\DeclareMathOperator*{\argmin}{arg\,min}

\newcommand{\be}{\begin{enumerate}}
\newcommand{\ee}{\end{enumerate}}
\newcommand{\bi}{\begin{itemize}}
\newcommand{\ei}{\end{itemize}}
\newcommand{\beq}{\begin{equation}}
\newcommand{\eeq}{\end{equation}}

\newcommand{\bp}{\begin{proof}}
\newcommand{\ep}{\end{proof}}
\newcommand{\bcor}{\begin{cor}}
\newcommand{\ecor}{\end{cor}}
\newcommand{\bthm}{\begin{thm}}
\newcommand{\ethm}{\end{thm}}
\newcommand{\blmm}{\begin{lmm}}
\newcommand{\elmm}{\end{lmm}}
\newcommand{\bdefn}{\begin{defn}}
\newcommand{\edefn}{\end{defn}}
\newcommand{\bprop}{\begin{prop}}
\newcommand{\eprop}{\end{prop}}
\newcommand{\bconj}{\begin{conj}}
\newcommand{\econj}{\end{conj}}
\newcommand{\bopm}{\begin{opm}}
\newcommand{\eopm}{\end{opm}}
\newcommand{\brmk}{\begin{rmk}}
\newcommand{\ermk}{\end{rmk}}

\newcommand{\suchthat}{\ | \ }

\newcommand{\mv}[1]{\mathbf{#1}}

\newtheorem{thm}{Theorem}[section]
\newtheorem{lmm}[thm]{Lemma}
\newtheorem{prop}[thm]{Proposition}
\newtheorem{cor}[thm]{Corollary}

\newdef{pbm}{Problem}
\newdef{question}{Question}
\newdef{assumption}{Assumption}
\newdef{opm}{Open Problem}
\newdef{conj}{Conjecture}
\newdef{ex}{Example}
\newdef{exer}{Exercise}
\newdef{defn}{Definition}
\newdef{alg}{Algorithm}
\newdef{rmk}{Remark}
\newdef{claim}{Claim}

\definecolor{Red}{RGB}{255,204,204}
\definecolor{Green}{RGB}{204,255,204}
\definecolor{Blue}{RGB}{204,204,255}

%% file: abstract.tex
\begin{abstract}
We define and study the {\bf F}unctional {\bf A}ggregate {\bf Q}uery ($\faq$)
problem, which captures common computational tasks across a very wide range
of domains including relational databases, logic, matrix and tensor computation, 
probabilistic graphical models, constraint satisfaction, and signal processing. 
Simply put, an $\faq$ is a declarative way of defining a new function from a 
database of input functions. 

We present $\InsideOut$, a dynamic programming algorithm, to evaluate an $\faq$.
The algorithm rewrites the input query into a set of
easier-to-compute $\faq$ sub-queries. Each sub-query is then evaluated using a
worst-case optimal relational join algorithm. The topic of designing algorithms 
to optimally evaluate the classic multiway join problem has seen exciting
developments in the past few years. Our framework tightly connects these new 
ideas in database theory with a vast number of application areas in a coherent
manner, showing potentially that -- with the right abstraction, blurring the 
distinction between data and computation -- a good 
database engine can be a general purpose constraint solver, relational data store,
graphical model inference engine, and matrix/tensor computation processor all at once.

The $\InsideOut$ algorithm is very simple, as shall be described in this paper.
Yet, in spite of solving an extremely general problem, its runtime either 
is as good as or improves upon the best known algorithm for the
applications that $\faq$ specializes to. These corollaries include computational
tasks in graphical model inference, matrix/tensor operations, relational joins, and 
logic. Better yet, $\InsideOut$ can be used within any database engine, because 
it is basically a principled way of rewriting queries. Indeed, it is 
already part of the $\LB$ database engine, helping efficiently answer 
 traditional database queries, graphical model inference queries, and train 
a large class of machine learning models inside the database itself.
\end{abstract}

%% file: intro.tex

\section{Introduction}


The following fundamental problems from diverse domains share
a common algebraic structure involving (generalized) sums of products.

\begin{ex}($\problemname{Matrix Chain Multiplication}$ ($\mcm$))
\label{ex:matrix mult}
Given a series of matrices $\mv A_1, \dots, \mv A_n$ over some field $\F$, 
where the dimension of $\mv A_i$ is $p_i \times p_{i+1}$, $i \in [n]$,
we wish to compute the product $\mv A = \mv A_1 \cdots \mv A_n$.
The problem can be reformulated as follows.
There are $n+1$ variables $X_1,\dots,X_{n+1}$ with domains 
$\Dom(X_i) = [p_i]$, for $i\in[n+1]$.
For $i \in [n]$, matrix $\mv A_i$ can be viewed as a function of two variables
\[ \psi_{i,i+1}: \Dom(X_i)\times \Dom(X_{i+1}) \to \F, \]
where $\psi_{i,i+1}(x,y) = (\mv A_i)_{xy}$.
The $\mcm$ problem is to compute the output function
\[ \varphi(x_1,x_{n+1}) =
   \sum_{x_2 \in\Dom(X_2)} \cdots \sum_{x_n \in \Dom(X_n)} \prod_{i=1}^n 
   \psi_{i,i+1}(x_i,x_{i+1}).
\]
\end{ex}

\begin{ex}(Maximum A Posteriori ($\map$) queries in 
      probabilistic graphical models ($\pgm$s))
   Consider a discrete graphical model represented by a hypergraph 
   $\calH = (\calV, \calE)$.
   There are $n$ discrete random variables $\calV = \{X_1,\dots,X_n\}$ on finite 
   domains $\Dom(X_i)$, $i\in[n]$, and $m=|\calE|$ {\em factors} 
   \[ \psi_S : \prod_{i \in S} \Dom(X_i) \to \R_+, \ S \in \calE. \]
   A typical inference task is to compute the marginal $\map$ estimates, 
   written in the form
   \[ \varphi(x_1,\dots,x_f) = 
      \max_{x_{f+1}\in\Dom(X_{f+1})} \cdots
      \max_{x_{n}\in\Dom(X_{n})} \prod_{S\in\calE}\psi_S(\mv x_S).
   \]
\end{ex}

\begin{ex}(Conjunctive query in RDBMS)
\label{ex:db:basic}
   Consider a schema with the following input relations:
   $R(a,b)$, 
   $S(b,c)$,
   $T(c,a)$, 
   where for simplicity let us say all attributes are integers.
   Consider the following query:
\begin{verbatim}
SELECT R.a
FROM R, S, T
WHERE R.b = S.b AND S.c = T.c AND T.a = R.a;
\end{verbatim}
The above query can be reformulated as follows.
Relation $R(a,b)$ is modeled by a function $\psi_R(a,b) \to
\{\true, \false\}$, where $\psi_R(a,b)=\true$ iff $(a,b) \in R$,
and relations $S(b,c)$ and $T(c, a)$ are modeled by similar functions $\psi_S(b, c), \psi_T(c, a)$. Now, computing the above query basically corresponds to computing the function $\varphi(a)\rightarrow\{\true,\false\}$, defined as:
\[\varphi(a)=
\bigvee_b \bigvee_c \psi_R(a, b)\wedge \psi_S(b, c)\wedge \psi_T(c, a).\]
\end{ex}

\begin{ex}($\problemname{\# Quantified Conjunctive Query}$ ($\sqcq$))
\label{ex:sqcq}
Let $\Phi$ be a first-order formula of the form
\[ \Phi(X_1,\dots,X_f) = \DarkBlue{Q_{f+1}} X_{f+1} \cdots 
   \DarkGreen{Q_n} X_n 
    \left( \bigwedge_{R\in\atoms(\Phi)} R \right), \]
where $Q_i \in \{\exists, \forall\}$, for $i>f$. 
The $\sqcq$ problem is to {\em count} the number of tuples in relation $\Phi$ 
on the free variables $X_1,\dots,X_f$.
To reformulate $\sqcq$,
construct a hypergraph $\calH = (\calV, \calE)$ as follows: $\calV$ is the set
of all variables $X_1,\dots,X_n$, and for each $R \in \atoms(\Phi)$
there is a hyperedge $S = \vars(R)$ consisting of all variables in $R$.
The atom $R$ can be viewed as a function indicating
whether an assignment $\mv x_S$ to its variables is satisfied by the atom;
namely $\psi_S(\mv x_S) = 1$ if $\mv x_S \in R$ and $0$ otherwise. 

For each $i \in \{f+1,\dots,n\}$ we define an aggregate operator
\[ \textstyle{\bigoplus^{(i)} = 
    \begin{cases}
    \max & \text{ if } Q_i = \exists, \\
    \times & \text{ if } Q_i = \forall.
    \end{cases}}
\]
Then, the $\sqcq$ problem above is to compute the {\em constant} function
\[ \varphi = \sum_{x_1\in\Dom(X_1)} \cdots \sum_{x_f\in\Dom(X_f)} 
   \mathop{\textstyle{\DarkBlue{\bigoplus^{(f+1)}}}}_{x_{f+1}\in\{0,1\}}
   \cdots
   \mathop{\textstyle{\DarkGreen{\bigoplus^{(n)}}}}_{x_{n}\in\{0,1\}}
   \prod_{S\in\calE}\psi_S(\mv x_S).
\]
\end{ex}

It turns out that these and dozens of other fundamental problems from 
constraint satisfaction ($\csp$), 
databases, 
matrix operations, 
$\pgm$ inference, 
logic,
coding theory,
and complexity theory
can be viewed as special instances
of a generic problem we call the 
{\bf F}unctional {\bf A}ggregate {\bf Q}uery, or
the $\faq$ problem, which we define
next. (See~\cite{faq-arxiv,AM00} for many more examples.)


Throughout the paper, we use the following convention.
Uppercase $X_i$ denotes a variable, and lowercase $x_i$ denotes a value in 
the domain $\Dom(X_i)$ of the variable. 
Furthermore, for any subset $S\subseteq [n]$, define
\begin{align*}
    \mv X_S &= (X_i)_{i\in S},
            & \mv x_S = (x_i)_{i\in S} \in \prod_{i\in S}\Dom(X_i).
\end{align*}
In particular, $\mv X_S$ is a tuple of variables and $\mv x_S$ is a
tuple of specific values with support $S$.
The input to $\faq$ is a set of functions and the output is a function
computed using a series of aggregates over the variables and
input functions.
More specifically, for each $i\in [n]$, let $X_i$ be a variable on some discrete 
domain $\Dom(X_i)$, where $|\Dom(X_i)|\ge 2$.
The $\faq$ problem is to compute the following function
\begin{equation}
\label{eq:gen-faq}
    \varphi(\mv x_{[f]}) =
    \mathop{\DarkBlue{\textstyle{\bigoplus^{(f+1)}}}}_{x_{f+1}\in\Dom(X_{f+1})}
       \cdots
    \mathop{\DarkGreen{\textstyle{\bigoplus^{(n)}}}}_{x_{n}\in\Dom(X_{n})}
    \mathop{\textstyle{\bigotimes}}_{S\in\calE}\psi_S(\mv x_S),
\end{equation}
where 
\bi
 \item $\calH=(\calV,\calE)$ is a multi-hypergraph. $\calV=[n]$ is the index
    set of the variables $X_i$, $i\in [n]$. Overloading notation, $\calV$ is
    also referred to as the set of variables.
 \item The set $F = [f]$ is the set of {\em free variables} for some 
    integer $0\le f\le n$. Variables in $\calV-F$ are called 
    {\em bound variables}. (Free and bound are logic terminologies. Free
    variables are group-by variables in database nomenclature.)
 \item $\D$ is a fixed domain, such as $\{0,1\}$, $\R^+$, $\Z$.
 \item For every hyperedge $S\in\calE$, 
    $\psi_S:\prod_{i\in S} \Dom(X_i) \to \D$ is an 
    {\em input} function (also called a {\em factor}).
 \item For every bound variable $i>f$, $\oplus^{(i)}$ is a binary (aggregate) 
    operator on the domain $\D$.
 \item And, for each bound variable $i>f$ either $\oplus^{(i)}=\otimes$ or 
    $(\D,\oplus^{(i)},\otimes)$ forms a commutative semiring \footnote{A triple 
   $(\D,\oplus,\otimes)$ is a {\em commutative semiring} if $\oplus$
and $\otimes$ are commutative binary operators over $\D$ satisfying the
following:
  (1) $(\D, \oplus)$ is a commutative monoid with an additive
                   identity, denoted by $\mv 0$.
  (2) $(\D, \otimes)$ is a commutative
monoid with a multiplicative
                   identity, denoted by $\mv 1$.
  (3) $\otimes$ distributes over $\oplus$.
  (4) For any element $e \in \mathbf D$, $e \otimes \mv 0
                       = \mv 0 \otimes e = \mv 0$.
} (with the same
    $\mv 0$ and $\mv 1$). Informally, this means that we can do addition and multiplication over $\D$ and still remain in the same set.

If $\oplus^{(i)} = \otimes$, then $\oplus^{(i)}$ is called a 
{\em product aggregate};
otherwise, it is a {\em semiring aggregate}. (We assume that there is at least one semiring aggregate.)
\ei
Because for $i > f$ every variable $X_i$ has its own aggregate $\bigoplus^{(i)}$ 
over all values $x_i \in \Dom(X_i)$, in the rest of the paper we will 
write $\bigoplus^{(i)}_{x_i}$ to mean 
$\displaystyle{\mathop{\textstyle{\bigoplus^{(i)}}}_{x_{i}\in\Dom(X_{i})}}$.

We will refer to $\varphi$ as an {\em $\faq$-query}.
We use $\faqcs$ to denote
the special case when there is a \emph{S}ingle \emph{S}emiring aggregate, 
i.e. $\oplus^{(i)}=\oplus, \forall i>f$, and $(\D, \oplus, \otimes)$ is a 
semiring~\cite{AM00}.

\begin{ex}(Aggregate query in RDBMS)
\label{ex:db}
   Consider the following query over relations
   $R(a,b)$, 
   $S(a,c)$,
   $T(b,c,d,e)$,
   $U(d,f)$,
   $V(e,f)$,
   $W(e,g)$,
   $Y(f,h)$, where all attributes are integers:
\begin{verbatim}
SELECT R.b, U.d, sum(W.e)
FROM R, S, T, U, V, W, Y
WHERE R.a = S.a AND R.b = T.b AND S.c = T.c 
      AND T.d = U.d AND T.e = V.e AND W.e = V.e 
      AND U.f = V.f AND Y.f = V.f GROUP BY R.b, U.d;
\end{verbatim}
We now explain how the above query can be reduced to an $\faq$
instance. Relation $R(a,b)$ is modeled with a function $\psi_R(a,b) \to
\{0,1\}$, where $\psi_R(a,b)=1$ iff $(a,b) \in R$. Similarly, we can think of
relations $S$, $T$, $U$, $V$, and $Y$ as functions 
$\psi_S$,
$\psi_T$,
$\psi_U$,
$\psi_V$,
$\psi_Y$,
with $\{0,1\}$ values.
We single out one relation $W(e,g)$ where the modeling is different:
$\psi_W(e,g) = e$ if $(e,g) \in W$ and $0$ otherwise.
The corresponding $\faq$-query is
\[ \varphi(b, d) = \sum_a \sum_c \sum_e \sum_f \sum_g \sum_h
   \psi_R
   \psi_S
   \psi_T
   \psi_U
   \psi_V
   \psi_W
   \psi_Y
\]
(For readability, we did not write the argument lists of the functions $\psi_R, \psi_S$, etc. They
should be obvious from context.) 
Note that a tuple in the output of the aggregate query has the schema $(b,d,\varphi(b,d))$.
The corresponding hypergraph is shown in
   Fig.~\ref{fig:example:graph}. The set of free variables is $F = \{b, d\}$.
   The domain is $\mv D = \Z$, the set of integers.
Note also that the above reduction to $\faq$ still works if we replace sum by another aggregate, e.g., max.
   
In order to explain later the connection of $\InsideOut$ to query rewriting, we also write the above query in $\LogiQL$, an extension of Datalog 
supported by the $\LB$ engine~\cite{LB}:
\begin{verbatim}
Q[b, d] = s <- agg<<s = total(e)>> R(a,b), S(a,c), 
    T(b,c,d,e), U(d,f), V(e,f), W(e,g), Y(f,h).
\end{verbatim}
In the above, {\tt agg} is short for aggregate, {\tt total} is equivalent to {\tt sum}
in {\tt SQL}, the notation {\tt Q[b,d]=s} means that the head predicate is
{\tt Q(b,d,s)} where {\tt (b,d)} is a key, hence the query computes 
{\tt Q(b,d,sum(e))}. 
\end{ex}

\input{fig_example}

The above example illustrates several important points.
First, when we defined the $\faq$ problem we did not specify how the input and
output factors are represented. The representation choice turns out to make a
huge difference in computational complexity~\cite{faq-arxiv}.
However, in practical applications the representation is usually the obvious one:
an input factor $\psi_S(\mv X_S)$ can be thought of as a table of tuples
$[\mv x_S, \psi_S(\mv x_S)]$, with the implicit assumption that if $\mv x_S$ is 
not in the table then its $\psi_S$-value is $\mv 0$. (This is the additive identity 
$\mv 0$ of the domain $\mv D$.)
Second, the reduction to $\faq$ is only at the syntax level. No real data conversion is 
necessary. All the data we need to obtain the functions $\psi_R$,
$\psi_T$ etc. are already in the input relations.
Third, in the mathematical definition of $\varphi(b, d)$ above, the domains of all 
variables are integers and so we have {\em infinite} sums. We could have 
restricted all variables to their active domains; but that is not necessary 
because summing over all integers or over the active domains give identical 
answer: tuples not present are assumed to have values $\mv 0$.

Now that we have established the scope of $\faq$, in the remainder of 
this paper we show a perhaps surprising result that an $\faq$ problem can be 
solved by {\em one} simple yet efficient algorithm. The algorithm can be implemented as a set
of ordinary database queries. The runtime matches or improves upon the best known 
runtimes in many application areas that the $\faq$ framework captures.
The runtime depends on the order of variable aggregates in the $\faq$ expression,
which naturally leads us to the question of how to re-order those aggregates to obtain the best runtime without changing the semantic meaning of the expression.

%% file: fig_example.tex

\pgfkeys{/tikz,
graph/vertex/.style={radius=.07},
graph/edge/.style={blue!70!black, line width=.7},
tree/node/.style={ellipse, draw=black},
tree/edge/.style={}}

\begin{figure}[h!]
{\centering
\subfloat[Query hypergraph]{
\begin{tikzpicture}[scale = .9, yscale=.8, every node/.style={transform shape}]

\coordinate (a) at (-1.25, 2.5);
\coordinate (b) at (0, 3);
\coordinate (c) at (0, 2);
\coordinate (d) at (0, 1);
\coordinate (e) at (0, 0);
\coordinate (f) at (1.25, .5);
\coordinate (g) at (-1, -1);
\coordinate (h) at (2, -.5);

\draw [graph/edge, green!80!black] (0, 1.5) ellipse (.4 and 2.3);
\draw [graph/edge, red] (a)--(b);
\draw [graph/edge, blue!80!black] (a)--(c);
\draw [graph/edge, cyan] (f)--(d);
\draw [graph/edge, orange] (f)--(e);
\draw [graph/edge, yellow!80!black] (f)--(h);
\draw [graph/edge, purple] (e)--(g);

\fill [graph/vertex] (b) circle node[right] {$b$};
\fill [graph/vertex] (c) circle node[right] {$c$};
\fill [graph/vertex] (d) circle node[left] {$d$};
\fill [graph/vertex] (e) circle node[left] {$e$};
\fill [graph/vertex] (a) circle node[left] {$a$};
\fill [graph/vertex] (f) circle node[right] {$f$};
\fill [graph/vertex] (h) circle node[right] {$h$};
\fill [graph/vertex] (g) circle node[left] {$g$};

\coordinate (ab) at ($(a)!.5!(b)$);
\coordinate (ac) at ($(a)!.5!(c)$);
\coordinate (fd) at ($(f)!.5!(d)$);
\coordinate (fe) at ($(f)!.5!(e)$);
\coordinate (fh) at ($(f)!.5!(h)$);
\coordinate (eg) at ($(e)!.5!(g)$);
\coordinate (bcde) at ($(c)!.5!(d)$);

\node [graph/edge, green!80!black] at (bcde) {$T$};
\node [graph/edge, above, red] at (ab) {$R$};
\node [graph/edge, below, blue!80!black] at (ac) {$S$};
\node [graph/edge, above, cyan] at (fd) {$U$};
\node [graph/edge, below, orange] at (fe) {$V$};
\node [graph/edge, below, yellow!80!black ] at (fh) {$Y$};
\node [graph/edge, below, purple] at (eg) {$W$};
\end{tikzpicture}
\label{fig:example:graph}
}
\subfloat[Tree decomposition]
{\begin{tikzpicture}[scale = .9, every node/.style={transform shape}]
\node (root) [tree/node] at (0, 0) {$b, c, d, e$};
\node (child1) [tree/node] at (-1.5, -1) {$a, b, c$};
\node (child2) [tree/node] at (1.5, -1) {$d,e,f$};
\node (child21) [tree/node] at (.5, -2) {$e, g$};
\node (child22) [tree/node] at (2.5, -2) {$f, h$};
\draw [tree/edge] (root)--(child1);
\draw [tree/edge] (root)--(child2);
\draw [tree/edge] (child2)--(child21);
\draw [tree/edge] (child2)--(child22);
\end{tikzpicture}
\label{fig:example:tree}
}
}
\caption{Query from Example~\ref{ex:db}}
\label{fig:running:example}
\end{figure}
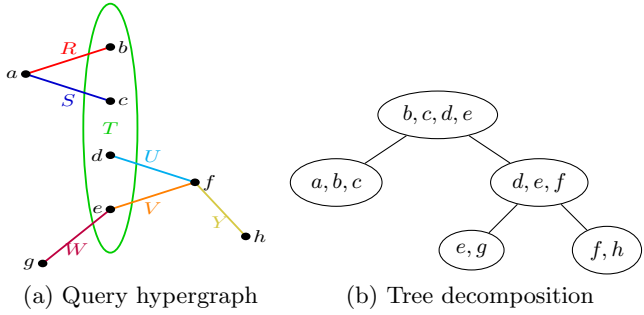

%% file: insideout.tex

\section{The {\large \sf InsideOut} Algorithm}
\label{sec:algo}

Parts of this section will be familiar to readers 
who have been exposed to elementary graphical models~\cite{MR2778120}. There 
are, however, a couple of ideas that are taken from new developments in
database theory~\cite{skew, NPRR12, faq}
 that are likely not known in the graphical model literature.
For each factor $\psi_S$, define its {\em size} to be the number of non-zero
points under its domain:
$ |\psi_S| := \left|\bigl\{ \mv x_S \suchthat 
                       \psi_S(\mv x_S) \neq \mv 0 
                     \bigr\}
               \right|.$

{\it \bf{Basic variable elimination.}}
To describe the intuition, we first explain $\InsideOut$ as it applies to
the special case of $\faqcs$ (or $\problemname{SumProd}$). 
The idea behind variable elimination 
\cite{DBLP:journals/ai/Dechter99,MR1426261,zhangpoole94} is to `fold' common
factors, exploiting the distributive law:
\begin{eqnarray*}
   && \bigoplus_{x_{f+1}}\cdots
   \bigoplus_{x_n}
   \bigotimes_{S\in\calE} \psi_S(\mv x_S)\\
   &=&\bigoplus_{x_{f+1}}\cdots
      \bigoplus_{x_{n-1}}
      \bigotimes_{S\in\calE-\partial(n)} \psi_S(\mv x_S)
      \otimes \underbrace{
         \left( \bigoplus_{x_n} \bigotimes_{S\in\partial(n)}
\psi_S(\mv x_S)\right)}_{\text{new factor } \psi_{U_n-\{n\}}},
\end{eqnarray*}
where the equality follows from the fact that $\otimes$ distributes over
$\oplus$, $\partial(n)$ denotes all edges incident to $n$ in $\calH$ 
and $U_n=\cup_{S\in\partial(n)} S$. 
Assume for now that we can somehow efficiently compute the intermediate
factor $\psi_{U_n-\{n\}}$. Then, the resulting problem is another instance of 
$\faqcs$ on a modified multi-hypergraph $\calH'$, 
constructed from $\calH$ by removing vertex $n$ along with
all edges in $\partial(n)$, and {\em adding back} a new hyperedge $U_n-\{n\}$.
Recursively, we continue this process until all variables $X_n,\dots,X_{f+1}$
are eliminated. Textbook treewidth-based results for $\pgm$ inference
are obtained this way \cite{MR2778120}.
In the database context (i.e.\ given an $\faq$-query over the Boolean semiring), the 
intermediate result $\psi_{U_n-\{n\}}$ is essentially an intermediate
relation of a query plan, the folding technique exploiting distributive law 
corresponds to ``pushing the aggregate down'' the query 
plan~\cite{Cohen:2006:UAF:1142473.1142480}.

{\it\bf Introducing the indicator projections.}
While correct, basic variable elimination as described above is potentially 
not very efficient for sparse input factors, i.e. factors where the number of non-zero
entries is much smaller than the product of the active domain sizes. 
This is because the product that was factored out of the scope of $X_n$
might annihilate many entries of the intermediate result
$\psi_{U_n-\{n\}}$, while we have spent so much time computing $\psi_{U_n-\{n\}}$.
For example, for an $S\notin\partial(n)$ such that $S\subseteq U_n$ and tuple $\mv y_S$ such that
$\psi_S(\mv y_S)=\mv 0$, we do not need to compute the entries
$\psi_{U_n-\{n\}}(\mv x_{U_n-\{n\}})$ for which 
$\mv y_{S} = \mv x_{S}$: those entries will be eliminated later 
anyhow.  The idea is then to only compute those 
$\psi_{U_n-\{n\}}(\mv x_{U_n-\{n\}})$ values that will ``survive'' the other 
factors later on. 
One simple way to achieve this would be to compute, for each 
$S \in \calE - \partial(n)$,  
an ``indicator factor'' that checks if $\psi_S(\mv x_S)$ is $\mv 0$ or not. 
Formally, for any two sets $T\subseteq S$, and a given factor $\psi_S$,
the function 
$\psi_{S/T} : \prod_{i \in T} \Dom(X_i) \to \mv D$
defined by
\[
   \psi_{S/T}(\mv x_T) :=
   \begin{cases}
      \mv 1 & \exists \mv x_{S-T} \text{ s.t. } 
          \psi_S(\mv x_T,\mv x_{S-T}) \neq \mv 0\\
      \mv 0 & \text{otherwise}
   \end{cases}
\]
is called the {\em indicator projection} of $\psi_S$ onto $T$.
Using indicator factors, $\InsideOut$ computes the following factor 
when marginalizing $X_n$ away:
\begin{multline}\label{eqn:sub:query}
    \psi_{U_n-\{n\}}(\mv x_{U_n-\{n\}}) =\\
    \bigoplus_{x_n} \left[ \left( \bigotimes_{S\in\partial(n)} \psi_S
      \right)
        \otimes \left( \bigotimes_{\substack{S\notin \partial(n),\\S\cap
            U_n\neq \emptyset}}
            \psi_{S/S\cap U_n}
            \right)\right].
\end{multline}
Another minor tweak is the observation that, if there is a hyperedge 
$S\in \calE-\partial(n)$ for which $S\subset U_n$, then we do not use 
the indicator projection $\psi_{S/S\cap U_n}$: we can use $\psi_S$ itself
to compute the intermediate factor $\psi_{U_n-\{n\}}$,
and then remove $\psi_S$ from $\calH'$.

\begin{ex}
We explain how the ideas above are implemented in 
Example~\ref{ex:db}. First, the 
order in which we choose to eliminate variables might have
a huge effect on the runtime. For now, let us assume that we somehow
decided to rewrite $\varphi(b,d)$ using the following variable order, where we
trace the first couple of steps of the $\InsideOut$ algorithm {\em without}
the indicator projection: (Example~\ref{ex:db:GYO} later explains how this order is related to the tree decomposition in Fig.~\ref{fig:example:tree}.)
\begin{eqnarray*}
 \varphi(b,d) &=& \sum_c \sum_a \sum_e \sum_f \sum_g \sum_h
   \psi_R
   \psi_S
   \psi_T
   \psi_U
   \psi_V
   \psi_W
   \psi_Y\\ 
     &=& \sum_c \sum_a \sum_e \sum_f \sum_g 
   \psi_R
   \psi_S
   \psi_T
   \psi_U
   \psi_V
   \psi_W
   \underbrace{\sum_h \psi_Y(f,h)}_{\psi_1(f)}\\
     &=& \sum_c \sum_a \sum_e \sum_f \sum_g 
   \psi_R
   \psi_S
   \psi_T
   \psi_U
   \psi_V
   \psi_W
   \psi_1 \\ 
     &=& \sum_c \sum_a \sum_e \sum_f 
   \psi_R
   \psi_S
   \psi_T
   \psi_U
   \psi_V
   \psi_1
   \underbrace{\sum_g \psi_W(e,g)}_{\psi_2(e)}\\
     &=& \sum_c \sum_a \sum_e \sum_f 
   \psi_R
   \psi_S
   \psi_T
   \psi_U
   \psi_V
   \psi_1
   \psi_2
\end{eqnarray*}
The first two steps are straightforward, where we eliminated $g$ and $h$.
In $\LogiQL$, these intermediate factors are computed with the following 
two rules
\begin{verbatim}
psi1[f] = s1 <- agg<<s1 = count()>> Y(f,h).
psi2[e] = s2 <- agg<<s2 = total(e)>> W(e,g).
\end{verbatim}
The mathematical abstraction corresponds to rewriting a query into a series of smaller queries.
Next, we explain how the indicator projection works when we eliminate variable
$f$. Out of the remaining factors
   $\psi_R$,
   $\psi_S$,
   $\psi_T$,
   $\psi_U$,
   $\psi_V$,
   $\psi_1$,
   and
   $\psi_2$,
   the following factors contain $f$: 
   $\psi_U(d,f)$ 
   $\psi_V(e,f)$ 
   and
   $\psi_1(f)$.
   If we were to multiply them together and marginalize away $f$, we would
   create a new factor $\psi_3(e,d) = \sum_f \psi_U\psi_V\psi_1$ over variables $\{e, d\}$.
   However, two other factors have variables that overlap with $\{e, d\}$, namely $\psi_T(b,c,d,e)$ and $\psi_2(e)$.
   For $\psi_T$, we include its indicator projection $\psi_{T/\{e,d\}}$
   in computing $\psi_3$.
   (We will see later in Example~\ref{ex:db:runtime} how including $\psi_{T/\{e,d\}}$ can actually speed up the computation of $\psi_3$ asymptotically.)
   For $\psi_2$, we can include $\psi_2$ itself.
  (Recall the minor tweak we mentioned above.)
  Overall,
   we end up with the following definition of $\psi_3$:
   \[ \psi_3(e,d) = \sum_f \psi_U \cdot \psi_V\cdot  \psi_1 \cdot \psi_2 \cdot  
      \psi_{T/\{e,d\}}. \]
   In $\LogiQL$, this sub-result is computed with two rules:
\begin{verbatim}
proj1(d,e) <- T(b,c,d,e). // projection rule
psi3[e,d] = s3 <- agg<<s3 = total(s1*s2)>> U(d,f), 
    V(e,f), psi1[f] = s1, psi2[e] = s2, proj1(d,e).
\end{verbatim}
After eliminating $f$, we are left with the following
\begin{eqnarray*}
   \varphi(b,d) &=& \sum_c\sum_a\sum_e 
   \psi_R
   \psi_S
   \psi_T
   \psi_3\\ 
   &=& \sum_c\sum_a
   \psi_R
   \psi_S
   \sum_e 
   \psi_3
   \psi_T\\ 
   &=& \sum_c\sum_a
   \psi_R
   \psi_S
   \underbrace{
   \sum_e 
   \psi_3
   \psi_T 
   \psi_{R/\{b\}} \psi_{S/\{c\}}}_{\psi_4(b,c,d)}
\end{eqnarray*}
leading to the following $\LogiQL$ rules
\begin{verbatim}
proj2(b) <- R(a,b).
proj3(c) <- S(a,c).
psi4[b,c,d] = s4 <- agg<<s4 = total(s3)>>
   psi3[e,d] = s3, T(b,c,d,e), proj2(b), proj3(c).
\end{verbatim}
At this point, we have 3 factors left $\psi_R(a,b)$,
$\psi_S(a,c)$, and $\psi_4(b,c,d)$. We eliminate $a$ then $c$ straightforwardly: 
   \[
   \sum_c\sum_a
   \psi_R
   \psi_S
   \psi_4\\ 
   =
   \sum_c
   \psi_4 
   \underbrace{
   \sum_a
   \psi_R
   \psi_S
   \psi_{4/\{b,c\}}
}_{\psi5(b,c)}\\
=
   \sum_c \psi_4 \psi_5.
   \]
Note that $\psi_{4/\{b, c\}}$ has values in $\{0, 1\}$ although $\psi_4$ can have any value in $\Z$.
The final $\LogiQL$ rules are
\begin{verbatim}
proj4(b,c) <- psi4[b,c,d] = s4. // indicator projection
psi5[b,c] = s5 <- agg<<s5 = count()>> 
                  R(a,b), S(a,c), proj4(b,c).
output[b,d] = t <- agg<<t = total(s4*s5)>> 
                  psi4[b,c,d] = s4, psi5[b,c] = s5.
\end{verbatim}
\label{ex:running:insideout}
\end{ex}

{\it\bf The general $\faq$ problem.}
The above strategy does not care if the variable aggregates where the same or
different: As long as $(\D, \oplus^{(n)}, \otimes)$ is a semiring, we can fold 
the common factors and eliminate $X_n$. Thus, $\InsideOut$ works almost as is 
for a general $\faq$ instance (as opposed to $\faqcs$).
Finally, when $\oplus^{(n)}=\otimes$ we simply swap the
two (identical) operators:
{\small
\begin{eqnarray*}
   \varphi(\mv x_{[f]})
   &=&
   \cdots 
   \mathop{\textstyle{\bigoplus^{(n-1)}}}_{x_{n-1}} 
   \mathop{\textstyle{\bigoplus^{(n)}}}_{x_n} 
   \mathop{\textstyle{\bigotimes}}_{S\in\calE} \psi_S(\mv x_S)\\
   &=&
   \cdots 
   \mathop{\textstyle{\bigoplus^{(n-1)}}}_{x_{n-1}} 
   \mathop{\textstyle{\bigotimes}}_{x_n \in \Dom(X_n)}
   \mathop{\textstyle{\bigotimes}}_{S\in\calE} \psi_S(\mv x_S)\\
   &=&
   \cdots 
   \mathop{\textstyle{\bigoplus^{(n-1)}}}_{x_{n-1}} 
   \mathop{\textstyle{\bigotimes}}_{S\in\calE} 
   \mathop{\textstyle{\bigotimes}}_{x_n \in \Dom(X_n)}
   \psi_S(\mv x_S)\\
   &=&
   \cdots 
   \mathop{\textstyle{\bigoplus^{(n-1)}}}_{x_{n-1}} 
   \mathop{\textstyle{\bigotimes}}_{S\notin\partial(n)} 
   \underbrace{\left(\psi_S(\mv x_S)\right)^{|\Dom(X_n)|}}_{\psi'_S}
   \mathop{\textstyle{\bigotimes}}_{S\in\partial(n)} 
   \underbrace{\mathop{\textstyle{\bigotimes}}_{x_n}
   \psi_S(\mv x_S)}_{\psi_{S-\{n\}}}.
\end{eqnarray*}
}
We are left with an $\faq$-instance whose hypergraph is exactly
$\calH' = \calH-\{n\}$: the hypergraph obtained form $\calH$ by removing 
vertex $n$ from
 the vertex set and all incident hyperedges.
The sub-problems are of the form of
{\em product marginalizations} of individual factors 
$\psi_S$ for $S\in \partial(n)$,
each of which can be computed in linear time in $|\psi_S|$.
The product marginalization step is algorithmically
much easier because it does not create the intermediate factor 
$\psi_{U_n-\{n\}}$. 
As for $S\notin\partial(n)$, we replace $\psi_S$ by the power factor
$\psi'_{S}(\mv x_{S}) = \left(\psi_S(\mv x_S)\right)^{|\Dom(X_n)|},$
which can be done in linear time with a $\log|\Dom(X_n)|$ blowup using the
repeated squaring algorithm.
Note the key fact that this power is with respect to the product aggregate
$\otimes$. In most (if not all) applications of $\faq$, there is one additional
property: most of the time, $\otimes$ is an idempotent operator over the active domain.
For example, in the $\sqcq$ problem $\otimes$ is the usual product operator and
the domain that it aggregates over is $\{0,1\}$ (before there is a sum outside).
In this case, 
$\psi'_{S}(\mv x_{S}) = \left(\psi_S(\mv x_S)\right)^{|\Dom(X_n)|} =
   \psi_S(\mv x_S),$
and we do not need to spend the linear nor $\log$-blowup time.
For more details on product idempotence, see~\cite{faq-arxiv}.

{\it\bf $\faq$ sub-problems as natural joins.}
In the above we have explained how $\InsideOut$ breaks a big problem into
smaller problems. In the product marginalization case, the sub-problems are
easy to solve: they can be solved in linear time. The most difficult problems,
however, are of the form~\eqref{eqn:sub:query}. This is exactly an $\faq$-query
where we marginalize out only one variable, with the remaining variables free.
Zooming in, problem~\eqref{eqn:sub:query} is of the form
\[ \psi_{U_n-\{n\}}(\mv x_{U_n-\{n\}}) := 
\bigoplus_{x_n} \bigotimes_{F\in\calE_n}\psi_F, \]
where $\calH_n=(U_n,\calE_n)$ is the sub-$\faq$-query hypergraph.
The problem is solved by computing 
$\psi_{U_n}(\mv x_{U_n}) := \bigotimes_{F\in\calE_n}\psi_F$ first.
Once the $\psi_{U_n}$ is computed, marginalizing away $X_n$ 
to obtain $\psi_{U_n-\{n\}}$ is trivial.

Computing the inner product is a natural join problem in disguise. Each 
input factor $\psi_S$ 
is represented using a table of tuples of the form 
$[\mv x_S, \psi_S(\mv x_S)]$.
Essentially, $\mv x_S$ is the (compound)
key and $\psi_S(\mv x_S)$ is the value in this relation.
Again, recall that entries not in the table have $\psi_S$-value $\mv 0$.
Hence, to compute $\psi_{U_n}$ we can first join the tables $\psi_S$
using only the key space. 
For each tuple $\mv x_{U_n}$ in the result of this join, we record the value
$\psi_{U_n}(\mv x) = \prod_{F\in\calE_n} \psi_{F}(\mv x_F)$.
The runtime is dominated by the natural join's runtime.

{\it\bf Worst-case optimal join algorithms.}
Computing the natural join is a very well-studied problem with exciting new
developments in the past decade or so.
There are new {\em worst-case optimal} algorithms~\cite{LFTJ,NPRR12,skew,anrr} 
that operate quite differently from traditional query plans, in the sense 
that they no longer compute one pairwise join at a
time, but instead process the query globally.  
While the vast majority of database engines today still rely on
traditional query plans, new, complex data analytics engines
are switching to worst-case optimal algorithms: $\LB$'s
engine~\cite{LB} is built on a worst-case optimal algorithm
called {\sf LeapFrog Triejoin}~\cite{LFTJ} ($\lftj$), and the {\em Myria}
data analytics platform supports a variant of
$\lftj$~\cite{DBLP:conf/sigmod/ChuBS15}.

We briefly outline these results here.
The generic form of the natural join problem can be posed in our hypergraph
language as
$Q = \ \Join_{F\in \calE} R_F$, where $\calH=(\calV,\calE)$ is the query
hypergraph. 
The vertices of this hypergraph consist of all attributes.
Each hyperedge $F\in\calE$ corresponds to an input relation $R_F$ 
whose attributes are $F$.
The natural join problem can be thought of as a constraint satisfaction 
problem: each input relation $R_F$ imposes a constraint where a tuple 
$\mv x_F$ satisfies the constraint if $\mv x_F \in R_F$.
A tuple $\mv x$ on all variables $\calV$ is an output of the join if 
the projection $\mv x_F$ satisfies $R_F$ for all $F\in\calE$.

$\lftj$~\cite{LFTJ} can be viewed as {\em backtracking-search} algorithm,
which was known some 50 years ago in the AI and constraint programming 
world~\cite{DBLP:journals/cacm/DavisLL62,DBLP:dblp_journals/jacm/GolombB65}.
(In contrast, by saving intermediate results $\psi_{U_n-\{n\}}$ instead of re-computing them each time,
$\InsideOut$ can be thought of as \emph{dynamic programming}.
The duality between
backtracking search and dynamic programming is 
well-known~\cite{Rossi:2006:HCP:1207782}.)
$\lftj$ fixes
some variable ordering $X_1,\dots,X_n$ of the query $Q$, then
performs ``leap-frogging'' to find the first binding $x_1$ that does not 
yet violate any constraints $R_F$; once $x_1$ is found, it looks for the first binding $x_2$ such that the partial tuple $(x_1,x_2)$ does not violate any 
constraint.  The algorithm proceeds this way until either a full binding
$\mv x$ is constructed in which case $\mv x$ is an output, or no good binding 
is found. For example, if no feasible binding for $x_3$ is found, then the 
algorithm backtracks to the next good binding of $x_2$.

The first advantage of backtracking search is that it requires only 
$O(1)$-extra space: it does not cache any computation. The second advantage, 
amazingly, is that a join algorithm based on back-tracking search such as 
$\lftj$ or others in~\cite{NPRR12, skew} are worst-case optimal, in the sense 
that the algorithm runs in time bounded by the worst-case output size.
To state the output size bound, we need the following notion. Define the
{\em fractional edge cover polytope} $\FECP(\calH)$ associated with a hypergraph
$\calH$ to be the set of all
vectors $\vec\lambda=(\lambda_F)_{F\in\calE}$ satisfying the following linear
constraints:
\[
   \vec\lambda \geq \mv 0, \text{ and }
   \sum_{F \in \calE, v \in F} \lambda_F \geq 1, \ \forall v \in \calV.
\]
A vector $\vec\lambda \in \FECP(\calH)$ is called a 
{\em fractional edge cover} of $\calH$.
The join output size is bounded above by $\prod_{F\in\calE}|R_F|^{\lambda_F}$,
for {\em any} $\vec\lambda \in \FECP(\calH)$.
The best bound $\agm(\calH)$, known as the $\agm$-bound~\cite{AGM08,GM06}, 
is obtained by solving the linear program
\begin{equation}
\min \Bigl\{ \sum_{F\in\calE}\lambda_F \log_2|R_F| : \vec\lambda\in
\FECP(\calH)\Bigr\}.
\end{equation}

\begin{ex}
\label{ex:db:runtime}
Consider the query computing $\psi_3$ in 
Example~\ref{ex:running:insideout}. The join query on the keys
has the following shape:
$Q = U(d,f) \Join V(e,f) \Join I(f) \Join J(e) \Join K(d,e).$
Then,
$\agm(Q) = 
|U|^{\lambda_{d,f}}
|V|^{\lambda_{e,f}}
|I|^{\lambda_f}
|J|^{\lambda_e}
|K|^{\lambda_{d,e}},
$
where $\vec\lambda$ is a fractional edge cover of the query's hypergraph.
Suppose all input relations have the same size $N$, then the optimal bound 
is obtained by setting $\lambda_{d,f}=\lambda_{d,e}=\lambda_{e,f}=1/2$,
and $\lambda_d=\lambda_e=0$. Worst-case optimal algorithms run in time
$\tilde O(N^{3/2})$ for this instance. Any traditional join-tree based 
plan runs in $\Omega(N^2)$-time for some input~\cite{NPRR12}.
Moreover, without the indicator projection of $T(b, c, d, e)$, there would be no $K(d, e)$ above, the best edge cover would be $\lambda_{d, f}=\lambda_{e, f}=1$, and the runtime would become $\Omega(N^2)$.
\end{ex}

{\it\bf Runtime analysis.}
Let $N$ denote the input size, $\outputsize$ the output size, and $K$ the 
set of $k \in [n]$ for which $\oplus^{(k)} \neq \otimes$ (note that $[f]
\subseteq K$).
Also, let $\agm(Q_k)$ denote the $\agm$-bound on the $k$th sub-query's
hypergraph $\calH_k$. Then, it is not hard to show~\cite{faq-arxiv} that the runtime 
of $\InsideOut$ is
\begin{equation}
   \tilde O\Bigl(
N + \sum_{k \in K}\agm(\calH_k)+\outputsize
\Bigr).
   \label{eqn:runtime1}
\end{equation}
The first term is input-preprocessing time, second
is the total subproblem solving time, and third is the unavoidable output
reporting time.
From~\eqref{eqn:runtime1}, 
we can write down a precise expression for the runtime of
$\InsideOut$.
Minimizing the resulting (somewhat complicated) expression leads to the dynamic
programming algorithm for the $\mcm$ problem 
and the $\problemname{FFT}$ algorithm for 
the $\problemname{DFT}$ (see~\cite{faq-arxiv} for details).

In the above discussion, we assumed that variables were eliminated in order 
$X_n,X_{n-1},\dots,X_{1}$. However, there is no reason to force
$\InsideOut$ to follow this particular order. 
In particular, there might be a different variable ordering for which
expression~\eqref{eqn:runtime1} is {\em a lot} smaller and the algorithm still
works correctly on that ordering (see~\cite{faq-arxiv}).
This is where the main technical contributions of our work in~\cite{faq-arxiv} begin.
We need to answer the following two fundamental questions:

{\it Question 1.}
How do we know which variable orderings are equivalent to the original
$\faq$-query expression? 

{\it Question 2.}
How do we find the ``best'' variable ordering among all
equivalent variable orderings? 

In the next two sections, we sketch how we answered the above two
questions and followup questions in theory and in practice.

%% file: theory.tex

\section{Theoretical contributions}

\input{contributions}
To answer the above questions, we start with some definitions.
A variable ordering $\sigma$ is {\em $\varphi$-equivalent} iff
permuting the variable aggregates of $\varphi$ using $\sigma$ gives an expression $\varphi'$
that is \emph{semantically-equivalent} to $\varphi$,
i.e.\ that always returns the same output as $\varphi$ {\em no matter} what the input is.
Let $\EVO(\varphi)$ denote the set of all $\varphi$-equivalent variable
orderings.

\begin{ex}
\label{ex:EVO}
The $\faq$ query $\varphi'$ below is $\varphi$-equivalent.
\begin{eqnarray*}
\varphi &=& \sum_a\sum_d \max_b \sum_c \psi_1(a, b) \psi_2(a, c) \psi_3 (c, d),\\
\varphi' &=& \sum_a\sum_c \sum_d \max_b \psi_1(a, b) \psi_2(a, c) \psi_3 (c, d).
\end{eqnarray*}
This is because $\varphi$ can be written as
\[\varphi = \sum_a\left[\left(\sum_d \sum_c \psi_2(a, c) \psi_3 (c, d)\right)
\left(\max_b  \psi_1(a, b)\right)\right].\]
\end{ex}

Now, for any $\sigma \in \EVO(\varphi)$, let $\calH^\sigma_k$ denote the $k$th
sub-query's hypergraph when we run $\InsideOut$ on $\sigma$. Ideally, we would
like to find $\sigma$ minimizing the expression
$\sum_{k \in K} \agm(\calH^\sigma_k)$. However, this expression is
data-dependent and thus it is a bit difficult to handle in a
mathematically clean way. We simplify our objective by approximating the
bound~\eqref{eqn:runtime1} a little:
we upperbound $\agm(\calH^\sigma_k)$ by the {\em fractional edge cover number} of the 
subgraph $\calH^\sigma_k$, 
i.e.  $\agm(\calH^\sigma_k) \leq N^{\rho^*(\calH^\sigma_k)}$, where
$\rho^*(\calH^\sigma_k) := \min\{ \sum_{F\in \calE}\lambda_F :
\vec\lambda\in\FECP(\calH^\sigma_k)\}$.
Then, \eqref{eqn:runtime1} is upperbounded by
$\tilde O\left(
      N^{\max_{k\in K} \rho^*(\calH^\sigma_k)}+\outputsize
   \right)$;
$\InsideOut$ on variable ordering $\sigma$ runs in $\tilde
O(N^{\faqw(\sigma)}+\outputsize)$-time, where
$\faqw(\sigma) := \max_{k\in K} \rho^*(\calH^\sigma_k)$.
Thus, to have the best runtime we would like to select an equivalent
ordering $\sigma$ with the smallest exponent $\faqw(\sigma)$, called the
{\em $\faq$-width} of an $\faq$-query:
\begin{equation}
    \faqw(\varphi) :=
    \min \left\{ \faqw(\sigma) \suchthat \sigma \in \EVO(\varphi) \right\} 
    \label{eqn:faqw}
\end{equation}
\begin{ex}
In Example~\ref{ex:EVO}, the original order in $\varphi$ has an $\faq$-width of 2, because eliminating $c$ first corresponds to joining $\psi_2$ and $\psi_3$ in time $\Omega(N^2)$.
In contrast, the order in $\varphi'$ has an $\faqw$ of 1, allowing to evaluate $\varphi$ in time $O(N)$.
\end{ex}
To solve the optimization problem~\eqref{eqn:faqw}, the first problem we have to
address is to precisely characterize the set $\EVO(\varphi)$.
Our approach, sketched in Fig.~\ref{fig:schematic}, is to construct an
{\em expression tree} of the $\faq$ query $\varphi$. The expression tree defines
a partially ordered set on the variables called the {\em precedence poset}.
Then, to complete the characterization of $\EVO$, we show that every ordering in
$\EVO$ is {\em component-wise equivalent} ($\CWE$) to a linear extension of $\LE(P)$. (See~\cite{faq-arxiv} for details.)
Thus, if we do not care about query complexity, we can take the orange path in
Fig~\ref{fig:schematic} and bruteforcedly compute an optimal variable ordering
$\sigma^*$, run it through $\InsideOut$, for a total runtime of 
$\tilde O(N^{\faqw(\varphi)}+\outputsize)$.

However, in some $\faq$-framework's applications such as in graphical models, we
cannot simply sweep query complexity under the rug. Moreover, computing the 
$\faqw$ is $\np$-hard because $\faqw$ is a strict generalization of
the fractional hypertree width ($\fhtw$),
which is $\np$-hard~\cite{2016arXiv161101090F} to compute. Hence, we 
find a good approximation for the $\faqw$. This is the green path in 
Fig.~\ref{fig:schematic}: from the expression tree, we construct a tree 
decomposition for $\calH$; then, from the GYO-elimination order of this tree 
decomposition we obtain a variable ordering $\bar \sigma$ for which we can
show that $\faqw(\bar \sigma) \leq \faqw(\sigma^*) + g(\faqw(\sigma^*))$, where
$g$ is any known approximation of $\fhtw$ 
(the best of which is due to Marx~\cite{Marx:2010:AFH:1721837.1721845}).

\begin{ex}
\label{ex:db:GYO}
The variable ordering used earlier in Example~\ref{ex:running:insideout}
is a GYO-elimination order for the tree decomposition in Fig.~\ref{fig:example:tree}.
(In GYO, when we eliminate variable $X_n$, the set $U_n$ becomes a bag of the tree decomposition whose children are the bags corresponding to $\partial(n)$.)
In particular, bags $\{f, h\}$, $\{e, g\}$, and $\{d, e, f\}$ resulted from eliminating $h, g,$ and $f$ respectively.
The tree decomposition has width $\fhtw=3/2$, same as the $\faqw$ of the variable ordering.
\end{ex}

\input{results}

From these ideas, we obtained many corollaries, some of which are summarized 
in Table~\ref{tab:results}. 
The results in Table~\ref{tab:results} span three areas: 
(1) $\csp$s and Logic; 
(2) $\pgm$s and 
(3) Matrix operations. 
Except for joins, problems in area (1) need the full generality of $\faq$ formulation, where
$\InsideOut$ either improves upon existing results or yields new results.
Problems in area (2) can already be reduced to $\faqcs$. Here, $\InsideOut$
improves upon known results since it takes advantage of Grohe and Marx's more
recent fractional hypertree width bounds. 
Finally, problems in area (3) of Table~\ref{tab:results} are classic.
$\InsideOut$ does not yield anything new here, but it is intriguing to be able
to explain the textbook dynamic programming algorithm for 
$\problemname{Matrix-Chain Multiplication}$~\cite{MR2002e:68001} 
as an algorithm to find a good
variable ordering for the corresponding $\faq$-instance.
The $\problemname{DFT}$ result is a re-writing of Aji and McEliece's
observation \cite{AM00}.

%% file: contributions.tex

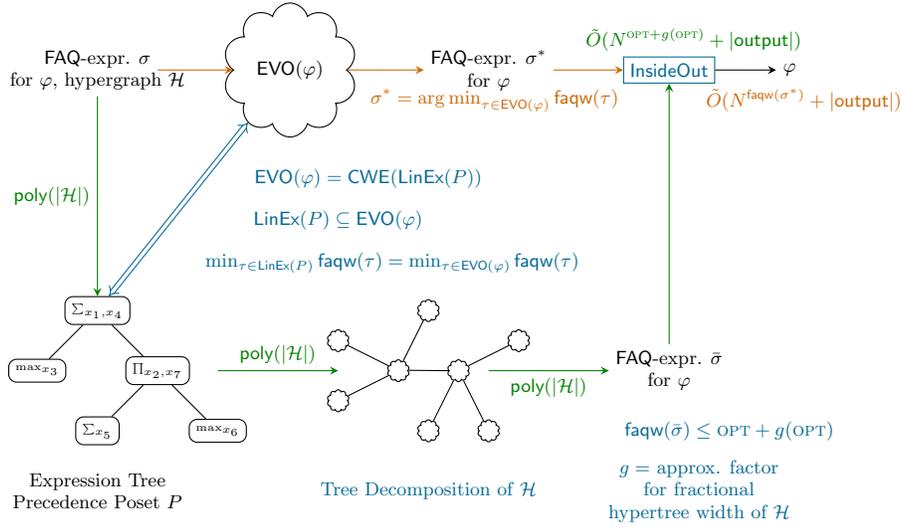
\begin{figure*}[!th]
\centering{
   \tikzset{>=stealth}
   \begin{tikzpicture}
      [every text node part/.style={align=center},
       et node/.style={rectangle, draw,font=\tiny,rounded corners=3pt}]

\begin{scope}[scale=0.8, every node/.style={scale=0.8}]
   \coordinate (input) at (0,6);
   \coordinate (et) at (1,1);
   \coordinate (td1) at (3,1);
   \coordinate (td2) at (5.5,1);

{
   \node[] (sigma) at (-1,6) {$\faq$-expr. $\sigma$ \\ for $\varphi$, hypergraph $\calH$};
}

{
   \node[draw,cloud] at (2.2,6) (evo) {$\EVO(\varphi)$};
   \path[->,DarkOrange] (input) edge (evo);
}
{
}

{
   \node[] at (5.5,6) (sigmastar) {$\faq$-expr. $\sigma^*$ \\ for $\varphi$};
   \path[->,DarkOrange] (evo) edge (sigmastar);
}
{
   \node[color=DarkOrange] at (5.6, 5.5) {$\sigma^* = \argmin_{\tau \in \EVO(\varphi)} \faqw(\tau)$};
}

{
   \node[draw,rectangle,color=MidnightBlue] at (8.5,6) (algo) {$\InsideOut$};
   \path[->,DarkOrange] (sigmastar) edge (algo);
}
{
   \node[color=DarkOrange,anchor=west] at (9, 5.5) {$\tilde O(N^{\faqw(\sigma^*)} + \outputsize)$};
}

{
   \node[] at (10.5,6) (phi) {$\varphi$};
   \path[->] (algo) edge (phi);
}

{
   \draw (-1,2) node[et node] (a) {$\sum_{x_1,x_4}$};
   \draw (-2,1) node[et node] (b) {$\max_{x_3}$};
   \draw (0,1) node[et node] (c) {$\prod_{x_2,x_7}$};
   \draw (-1,0) node[et node] (d) {$\sum_{x_5}$};
   \draw (1,0) node[et node] (e) {$\max_{x_6}$};
   \draw (a) -- (b);
   \draw (a) -- (c);
   \draw (c) -- (d);
   \draw (c) -- (e);
   \path[->,DarkGreen] (sigma) edge node[left] {{\sf poly}($|\calH|$)} (a);
   \node[] at (-1,-1) {Expression Tree \\ Precedence Poset $P$};
}

{
   \draw[implies-implies,color=MidnightBlue,double equal sign distance] (evo) -- (a);
   \node[color=MidnightBlue] at (3.5,4.2) {$\EVO(\varphi) = \CWE(\LE(P))$};
   \node[color=MidnightBlue] at (3.0,3.5) {$\LE(P) \subseteq \EVO(\varphi)$};
}

{
   \node[color=MidnightBlue] at (3.9,2.8) {$\min_{\tau \in \LE(P)}\faqw(\tau) = \min_{\tau \in \EVO(\varphi)} \faqw(\tau)$};
}

{
   \path[->,DarkGreen] (et) edge node[above] {{\sf poly}$(|\calH|)$} (td1);
   \node[color=MidnightBlue] (td) at (4.5,-1) {Tree Decomposition of $\calH$};
   \node[cloud,draw] (1) at (4,1) {};
   \node[cloud,draw] (2) at (4.5,2) {};
   \node[cloud,draw] (3) at (3,1.5) {};
   \node[cloud,draw] (4) at (3,0.5) {};
   \node[cloud,draw] (5) at (5,1) {};
   \node[cloud,draw] (6) at (4.5,0) {};
   \node[cloud,draw] (7) at (5.5,0) {};
   \node[cloud,draw] (8) at (6,1.5) {};
   \draw (1) -- (2);
   \draw (1) -- (3);
   \draw (1) -- (4);
   \draw (1) -- (5);
   \draw (5) -- (6);
   \draw (5) -- (7);
   \draw (5) -- (8);
}

{
   \node[] at (8.5,1) (sigmabar) {$\faq$-expr. $\bar \sigma$ \\ for $\varphi$};
   \path[->,DarkGreen] (td2) edge node[below] {{\sf poly}$(|\calH|)$} (sigmabar);
   \node[color=MidnightBlue] at (9.5,0) {$\faqw(\bar \sigma) \leq \opt+g(\opt)$};
   \node[color=MidnightBlue] at (9,-1) {$g = $ approx. factor \\ for
   fractional \\ hypertree width of $\calH$};
}

{
   \path[->,DarkGreen] (sigmabar) edge (algo);
   \node[color=DarkGreen,anchor=west] at (7, 6.5) {$\tilde O(N^{\opt+g(\opt)} + \outputsize)$};
}
\end{scope}

\end{tikzpicture}
}
\caption{Sketch of main technical contributions} 
\label{fig:schematic}
\end{figure*}

%% file: results.tex

\begin{table*}[th!]
{\small
\centering
{\renewcommand{\arraystretch}{1.5}
\begin{tabular}{|l|l|c|c|}
\hline
Problem & $\faq$ formulation & Previous Algo. & {\bf Our Algo.}\\
\hline
\rowcolor{Red}
$\sqcq$ & $\displaystyle{\sum_{(x_1,\dots,x_f)}} 
        \textstyle{\bigoplus^{(f+1)}_{x_{f+1}}\cdots\bigoplus^{(n)}_{x_n}}
        \displaystyle{\prod_{S\in\calE}\psi_S(\mv x_S)}$
        & No non-trivial algo & $\tilde
        O(N^{\faqw(\varphi)}+Z)$\\
\rowcolor{Red}
& ~~~~~~~ where $\textstyle{\bigoplus^{(i)}}\in\{\max,\times\}$  & & \\
\rowcolor{Red}
$\qcq$ &  $\textstyle{\bigoplus^{(f+1)}_{x_{f+1}}\cdots\bigoplus^{(n)}_{x_n}}
           \displaystyle{\prod_{S\in\calE}\psi_S(\mv x_S)}$
        & $\tilde O(N^{\problemname{PW}(\calH)}+Z)$~\cite{DBLP:conf/lics/ChenD12}&$\tilde O(N^{\faqw(\varphi)}+Z)$\\
\rowcolor{Red}
& ~~~~~~~ where $\textstyle{\bigoplus^{(i)}}\in\{\max,\times\}$ & & \\
\rowcolor{Red}
$\scq$ & $\displaystyle{\sum_{(x_1,\dots,x_f)} 
\max_{x_{f+1}}\cdots\max_{x_n}
        \prod_{S\in\calE}\psi_S(\mv x_S)}$
        & $\tilde
        O(N^{\problemname{DM}(\calH)}+Z)$~\cite{DBLP:conf/icdt/DurandM13}&
        $\tilde O(N^{\faqw(\varphi)}+Z)$\\
\rowcolor{Red}
Joins  & $\bigcup_{\mv x}\bigcap_{S\in\calE} \psi_S(\mv x_S)$
        & $\tilde{O}\left( N^{\fhtw(\calH)}+Z\right)$~\cite{GM06} & $\tilde{O}\left( N^{\fhtw(\calH)}+Z\right)$\\
\rowcolor{Green}
Marginal Distribution & $\displaystyle{\sum_{(x_{f+1},\dots,x_n)} \prod_{S\in\calE} \psi_S(\mv
x_S)}$  & $\tilde O(N^{\htw(\varphi)}+Z)$~\cite{KDLD05}
&$\tilde O(N^{\faqw(\varphi)}+Z)$\\
\rowcolor{Green}
MAP query & $\displaystyle{\max_{(x_{f+1},\dots,x_n)} \prod_{S\in\calE}
\psi_S(\mv x_S)}$ &
$\tilde O(N^{\htw(\varphi)}+Z)$~\cite{KDLD05}
& $\tilde O(N^{\faqw(\varphi)}+Z)$\\
\rowcolor{Blue}
Matrix Chain Mult. & $\displaystyle{\sum_{x_2,\dots,x_{n}} \prod_{i=1}^{n}
\psi_{i,i+1}(x_i,x_{i+1})}$ & DP bound~\cite{MR2002e:68001}  & DP bound\\
\rowcolor{Blue}
DFT & $\displaystyle{\sum_{(y_0,\dots,y_{m-1})\in\Z_p^m}b_y\cdot \prod_{0\le
j+k<m} e^{i2\pi\frac{x_j\cdot y_k}{p^{m-j-k}}}}$& $O(N\log_p{N})$~\cite{fft}& $O(N\log_p{N})$\\
\hline
\end{tabular}
}
\caption{Runtimes of algorithms assuming optimal variable ordering is given.
    Problems shaded red are in CSPs and logic ($\D=\{0,1\}$  for CSP and $\D=\N$ for \#CSP), problems
shaded green fall under PGMs ($\D=\R_+$), and problems shaded blue 
fall under matrix operations ($\D=\mathbb{C}$). 
$N$ denotes the size of the largest factor (assuming they are represented with
the listing format). 
$\htw(\varphi)$ is the notion of integral cover width 
for $\pgm$.
$\problemname{PW}(\calH)$ is the {\em optimal width of a prefix graph} of 
$\calH$ 
and 
$\problemname{DM}(\calH) = \poly(\Fss(\calH), \fhtw(\calH))$, where 
$\Fss(\calH)$ is the $[f]$-quantified star size. 
$Z$ is the output size in listing representation.
%
Our width $\faqw(\varphi)$ is never worse than any of the three and
there are classes of queries where ours is unboundedly better than all three.
In DFT, $N=p^m$ is the length of the input vector.
$\tilde O$ hides a factor of $\poly(|\calH|)\cdot \log N$.}
\label{tab:results}
}
\end{table*}

%% file: practice.tex

\section{Practical implications}

In this section we address two questions the readers might have regarding
$\InsideOut$: (1) hurdles one might face in a practical implementation of
$\InsideOut$, and (2) whether practical implications are as good as what the
theory says.

\paragraph*{Additional hurdles and how to solve them}
There are a couple of problems we have to solve to implement $\InsideOut$ 
effectively.

The first problem is, in real-world queries, we do not just have materialized predicates as inputs,
we also have predicates such as $a<b$, $a+b=c$, negations and so on. 
These predicates do not have a ``size.''
To solve this problem, one solution is to set the ``size'' of those 
predicates to be $\infty$
while computing the $\agm$-bound. For instance, if we have a sub-query of the form
$Q \leftarrow R(a,b),S(b,c),a+b=c$, where $R$ and $S$ are input
materialized predicates of size $N$, then by setting the size of $a+b=c$
to be infinite, $\agm(Q) = N^2$. This solution does
not work for two reasons. (1) If we knew $a+b=c$, then it is easy to infer 
that $|Q| \leq N$ and also to compute $Q$ in time $\tilde O(N)$:
scan over tuples in $R$, use $a+b=c$ to compute $c$, and see if $(b,c) \in S$.
In other words, the $\agm$-bound is no longer tight. 
(2) This solution may give an $\infty$-bound when the output size is clearly
bounded. Consider, for example, the query $Q \leftarrow R(a),S(b),a+b=c$; in 
this case, $\{a,b,c\}$ is the only hyperedge covering vertex $c$ in the
fractional edge cover.
Our implementation at $\LB$ makes use of generalizations of $\agm$ to 
queries with functional dependencies and immaterialized predicates (such as
$a+b=c$). These new bounds are based on a linear program 
whose variables are marginal entropies~\cite{DBLP:conf/pods/KhamisNS16,panda}.

The second problem is to select a good variable ordering to run $\InsideOut$
on. In principle, one does not have to use the $\agm$-bound or the
bounds from~\cite{DBLP:conf/pods/KhamisNS16,panda} to estimate the cost of
an $\faq$ subquery. If one were to implement $\InsideOut$ inside any RDBMS, one
could poll that RDBMS's optimizer to figure out the cost of a 
given variable ordering. However, there are $n!$ variable orderings, and 
optimizer's cost estimation is time-consuming. Furthermore, some subqueries
have inputs which are intermediate results. 
Hence, it is much faster to compute a variable ordering minimizing the
$\faqw$ of the query, defined on the bounds 
in~\cite{DBLP:conf/pods/KhamisNS16,panda}. As the problem is $\np$-hard, 
either an approximation algorithm~\cite{faq} or a greedy heuristic suffices
in our experience.

$\InsideOut$ is bottom-up dynamic programming. We can also solve $\faq$
queries with top-down (memoized) 
dynamic programming. In hindsight, this was the approach
that Bakibayev et al.~\cite{DBLP:journals/pvldb/BakibayevKOZ13} and 
Olteanu and Z\'avodn\'y~\cite{OZ15} took to solve $\faq$ over a single
semiring.
We can also limit the amount of memoization in a top-down strategy
to attain performance gain in some cases~\cite{KalinskyEK17}.

\paragraph*{Practical Impact}

It is trivial to construct classes of queries on real datasets for which 
$\InsideOut$-style of algorithms gives arbitrarily large speedups over 
traditional RDBMSs. In fact, even when dynamic programming does not take 
effect, the speedup of backtracking search (and thus
worst-case optimal algorithms) over traditional query plans is already 
huge~\cite{DBLP:conf/sigmod/NguyenABKNRR14}. 
The impact of the $\faq$-framework and the $\InsideOut$ algorithm, however,
go much beyond these toy queries (even when run on real datasets).

$\InsideOut$ is a component of $\LB$'s effort to extend $\LogiQL$ to be a 
probabilistic programming language~\cite{DBLP:conf/icdt/BaranyCKOV16}, as 
part of DARPA's PPAML and MUSE programs. The component the algorithm handles is 
inference in discrete graphical models.

Learning from the beautiful work of Olteanu and 
Schleich~\cite{DBLP:journals/pvldb/OlteanuS16, DBLP:conf/sigmod/SchleichOC16}, 
we realized~\cite{indblearn} 
that $\InsideOut$ 
can be used to train a large class of machine learning models {\em inside}
the database. Our implementation showed orders of magnitude speedup over the 
traditional data modeler route
of exporting the data and running it through \texttt{R} or \texttt{Python}.
These models are trained with different variations of gradient descents, 
whose (pre-)computation steps are $\faq$ queries. What is much more interesting
than the vanilla $\faq$ framework we presented above is that, in these 
applications, we want to compute {\em many} (in the 100K-range) $\faq$ queries all at 
once, making dynamic programming much more crucial to the performance.
Another related approach was considered in~\cite{Kumar:2015:LGL:2723372.2723713}.

%% file: conclusion.tex
\section{Concluding Remarks}

The $\faq$ framework showed that many common computational tasks over a very 
wide range of domains such as CSPs, machine learning, relational database,
logic, and matrix computations can be performed {\em inside} a database
using the same abstraction. The main idea is to blur the line between data
and computation, as we use the database to store, compute, and
process functions.
The glue of the framework is a simple dynamic programming algorithm called
$\InsideOut$, which can be cast as a query-rewriting method and thus it is
readily implementable within any RDBMS.
These ideas are implemented, tested, and validated within the $\LB$
database system. Our theory predicts practice very well, which is a 
beautiful thing to see.

%% file: main.bbl
\def\cprime{$'$}
\begin{thebibliography}{10}

\bibitem{anrr}
M.~{Abo Khamis}, H.~Q. Ngo, C.~R{\'e}, and A.~Rudra.
\newblock Joins via geometric resolutions: Worst-case and beyond.
\newblock In {\em PODS}, pages 213--228. ACM, 2015.

\bibitem{faq-arxiv}
M.~{Abo Khamis}, H.~Q. Ngo, and A.~Rudra.
\newblock {FAQ:} questions asked frequently.
\newblock {\em CoRR}, abs/1504.04044, 2015.

\bibitem{faq}
M.~{Abo Khamis}, H.~Q. Ngo, and A.~Rudra.
\newblock {FAQ:} questions asked frequently.
\newblock In {\em PODS}, 2016.

\bibitem{DBLP:conf/pods/KhamisNS16}
M.~{Abo Khamis}, H.~Q. Ngo, and D.~Suciu.
\newblock Computing join queries with functional dependencies.
\newblock In {\em PODS}, pages 327--342, 2016.

\bibitem{panda}
M.~{Abo Khamis}, H.~Q. {Ngo}, and D.~{Suciu}.
\newblock What do shannon-type inequalities, submodular width, and disjunctive
  datalog have to do with one another?
\newblock In {\em PODS}, 2017.

\bibitem{AM00}
S.~M. Aji and R.~J. McEliece.
\newblock The generalized distributive law.
\newblock {\em {IEEE} Transactions on Information Theory}, 46(2):325--343,
  2000.

\bibitem{LB}
M.~Aref, B.~ten Cate, T.~J. Green, B.~Kimelfeld, D.~Olteanu, E.~Pasalic, T.~L.
  Veldhuizen, and G.~Washburn.
\newblock Design and implementation of the {LogicBlox} system.
\newblock In {\em SIGMOD}, 2015.

\bibitem{AGM08}
A.~Atserias, M.~Grohe, and D.~Marx.
\newblock Size bounds and query plans for relational joins.
\newblock In {\em FOCS}, pages 739--748. IEEE Computer Society, 2008.

\bibitem{DBLP:journals/pvldb/BakibayevKOZ13}
N.~Bakibayev, T.~Kocisk{\'{y}}, D.~Olteanu, and J.~Zavodny.
\newblock Aggregation and ordering in factorised databases.
\newblock {\em {PVLDB}}, 6(14):1990--2001, 2013.

\bibitem{DBLP:conf/icdt/BaranyCKOV16}
V.~B{\'{a}}r{\'{a}}ny, B.~ten Cate, B.~Kimelfeld, D.~Olteanu, and Z.~Vagena.
\newblock Declarative probabilistic programming with datalog.
\newblock In {\em ICDT}, pages 7:1--7:19, 2016.

\bibitem{DBLP:conf/lics/ChenD12}
H.~Chen and V.~Dalmau.
\newblock Decomposing quantified conjunctive (or disjunctive) formulas.
\newblock In {\em LICS}, 2012.

\bibitem{DBLP:conf/sigmod/ChuBS15}
S.~Chu, M.~Balazinska, and D.~Suciu.
\newblock From theory to practice: Efficient join query evaluation in a
  parallel database system.
\newblock In {\em SIGMOD}, pages 63--78, 2015.

\bibitem{Cohen:2006:UAF:1142473.1142480}
S.~Cohen.
\newblock User-defined aggregate functions: Bridging theory and practice.
\newblock In {\em SIGMOD '06}, pages 49--60.

\bibitem{fft}
J.~W. Cooley and J.~W. Tukey.
\newblock {An algorithm for the machine calculation of complex Fourier series}.
\newblock {\em Mathematics of Computation}, 19:297--301, 1965.

\bibitem{MR2002e:68001}
T.~H. Cormen, C.~E. Leiserson, R.~L. Rivest, and C.~Stein.
\newblock {\em Introduction to algorithms}.
\newblock MIT Press, 2nd edition, 2001.

\bibitem{DBLP:journals/cacm/DavisLL62}
M.~Davis, G.~Logemann, and D.~W. Loveland.
\newblock A machine program for theorem-proving.
\newblock {\em Commun. {ACM}}, 5(7):394--397, 1962.

\bibitem{DBLP:journals/ai/Dechter99}
R.~Dechter.
\newblock Bucket elimination: {A} unifying framework for reasoning.
\newblock {\em Artif. Intell.}, 113(1-2):41--85, 1999.

\bibitem{DBLP:conf/icdt/DurandM13}
A.~Durand and S.~Mengel.
\newblock Structural tractability of counting of solutions to conjunctive
  queries.
\newblock In {\em ICDT}, pages 81--92, 2013.

\bibitem{2016arXiv161101090F}
W.~{Fischl}, G.~{Gottlob}, and R.~{Pichler}.
\newblock {General and Fractional Hypertree Decompositions: Hard and Easy
  Cases}.
\newblock {\em ArXiv e-prints}, Nov. 2016.

\bibitem{DBLP:dblp_journals/jacm/GolombB65}
S.~W. Golomb and L.~D. Baumert.
\newblock Backtrack programming.
\newblock {\em J. {ACM}}, 12(4):516--524, 1965.

\bibitem{GM06}
M.~Grohe and D.~Marx.
\newblock Constraint solving via fractional edge covers.
\newblock In {\em SODA}, pages 289--298, 2006.

\bibitem{KalinskyEK17}
O.~Kalinsky, Y.~Etsion, and B.~Kimelfeld.
\newblock Flexible caching in trie joins, 2017.
\newblock To appear in EDBT.

\bibitem{KDLD05}
K.~Kask, R.~Dechter, J.~Larrosa, and A.~Dechter.
\newblock Unifying tree decompositions for reasoning in graphical models.
\newblock {\em Artif. Intell.}, 166(1-2), 2005.

\bibitem{MR2778120}
D.~Koller and N.~Friedman.
\newblock {\em Probabilistic graphical models}.
\newblock Adaptive Computation and Machine Learning. MIT Press, 2009.
\newblock Principles and techniques.

\bibitem{Kumar:2015:LGL:2723372.2723713}
A.~Kumar, J.~Naughton, and J.~M. Patel.
\newblock Learning generalized linear models over normalized data.
\newblock In {\em SIGMOD}, pages 1969--1984. ACM, 2015.

\bibitem{Marx:2010:AFH:1721837.1721845}
D.~Marx.
\newblock Approximating fractional hypertree width.
\newblock {\em ACM Trans. Algorithms}, 6(2):29:1--29:17, Apr. 2010.

\bibitem{indblearn}
H.~Q. Ngo, X.~Nguyen, D.~Olteanu, and M.~Schleich.
\newblock In-database learning with sparse tensors, 2017.
\newblock Manuscript.

\bibitem{NPRR12}
H.~Q. Ngo, E.~Porat, C.~R{\'e}, and A.~Rudra.
\newblock Worst-case optimal join algorithms.
\newblock In {\em PODS}, pages 37--48, 2012.

\bibitem{skew}
H.~Q. Ngo, C.~R{\'e}, and A.~Rudra.
\newblock Skew strikes back: New developments in the theory of join algorithms.
\newblock In {\em SIGMOD RECORD}, pages 5--16, 2013.

\bibitem{DBLP:conf/sigmod/NguyenABKNRR14}
D.~T. Nguyen, M.~Aref, M.~Bravenboer, G.~Kollias, H.~Q. Ngo, C.~R{\'{e}}, and
  A.~Rudra.
\newblock Join processing for graph patterns: An old dog with new tricks.
\newblock In {\em GRADES}, pages 2:1--2:8, 2015.

\bibitem{DBLP:journals/pvldb/OlteanuS16}
D.~Olteanu and M.~Schleich.
\newblock {F:} regression models over factorized views.
\newblock {\em {PVLDB}}, 9(13):1573--1576, 2016.

\bibitem{OZ15}
D.~Olteanu and J.~Z\'{a}vodn\'{y}.
\newblock Size bounds for factorised representations of query results.
\newblock {\em ACM Trans. Datab. Syst.}, 40(1), 2015.

\bibitem{Rossi:2006:HCP:1207782}
F.~Rossi, P.~v. Beek, and T.~Walsh.
\newblock {\em Handbook of Constraint Programming (Foundations of Artificial
  Intelligence)}.
\newblock Elsevier Science Inc., 2006.

\bibitem{DBLP:conf/sigmod/SchleichOC16}
M.~Schleich, D.~Olteanu, and R.~Ciucanu.
\newblock Learning linear regression models over factorized joins.
\newblock In {\em SIGMOD}, pages 3--18, 2016.

\bibitem{LFTJ}
T.~L. Veldhuizen.
\newblock Triejoin: A simple, worst-case optimal join algorithm.
\newblock In {\em ICDT}, pages 96--106, 2014.

\bibitem{zhangpoole94}
N.~Zhang and D.~Poole.
\newblock {A simple approach to Bayesian network computations}.
\newblock In {\em Canadian AI}, pages 171--178, 1994.

\bibitem{MR1426261}
N.~L. Zhang and D.~Poole.
\newblock Exploiting causal independence in {B}ayesian network inference.
\newblock {\em J. Artificial Intelligence Res.}, 5:301--328, 1996.

\end{thebibliography}
